\documentclass[10pt, conference, compsocconf]{IEEEtran}
\usepackage{url}
\ifCLASSINFOpdf
  \usepackage[pdftex]{graphicx}
\else
\fi
\begin{document}

%

\title{Exploiting n-gram location for intrusion detection}


\author{\IEEEauthorblockN{Fabrizio Angiulli, Luciano Argento, Angelo Furfaro}
\IEEEauthorblockA{DIMES -- University of Calabria\\
P. Bucci, 41C\\
I-87036 Rende (CS), Italy\\
Email: \{f.angiulli, l.argento, a.furfaro\}@dimes.unical.it}
}

%


\maketitle

\begin{abstract}
Signature-based and protocol-based intrusion detection systems (IDS) are 
employed as means to reveal content-based network attacks. 
Such systems have proven to be effective in identifying known intrusion attempts
and exploits but they fail to recognize new types of attacks or 
carefully crafted variants of well known ones. This paper presents the design
and the development of an anomaly-based IDS technique 
which is able to detect content-based attacks carried out over application level
protocols, like HTTP and FTP. 
In order to identify anomalous packets, the payload is split up in chunks of
equal length and the n-gram technique is used to 
learn which byte sequences usually appear in each chunk. 
The devised technique builds a different model for each pair  $\langle$protocol of
interest, packet length$\rangle$  and 
uses them to classify the incoming traffic. Models are build by means of a
semi-supervised approach.
Experimental results witness that the technique achieves an excellent accuracy
with a very low  false positive rate.
\end{abstract}

\begin{IEEEkeywords}
Intrusion detection systems; Semi-supervised learning; N-grams; Anomaly detection; FTP traffic;

\end{IEEEkeywords}

%
\IEEEpeerreviewmaketitle


\section{Introduction}

Cyber security is a set of technologies, processes, means and practices designed
to protect computers, data, programs interconnected from unauthorized access or attacks. 
Network security is a crucial aspect of cyber security due to the
ubiquitous diffusion of the Internet. User authentication, data encryption, avoiding programming 
errors and firewalls are examples of protection techniques employed in 
order to improve security, but they generally lack the capabilities of protecting
against malicious mobile code and insider attacks. 
Therefore systems require an additional security layer, despite the prevention techniques,
and intrusion detection has proven to be an effective wall against a broad range of network attacks.
Intrusion detection systems monitor the activities in a computer system or network
and analyse them to recognise possible incidents, which are imminent threats or actual
violations of computer security policies or standard security practices. 

Intrusion detection is classified into two 
types: misuse intrusion detection and anomaly intrusion detection. Misuse
intrusion detection uses patterns, also known as signatures, that characterise 
known attacks, which exploit weaknesses in systems and application software, to
successfully identify malicious activities.
Anomaly intrusion detection aims at modelling the normal usage behavior to identify
signs of possible intrusions. 

There exist attacks which exploit some vulnerabilities of a service or
application by delivering a bad payload. It is possible to detect these attacks
by inspecting the packets payload. A lot of IDS use \textit{n}-grams for packets
analysis \cite{davis2011data},  some of such techniques are described in this section.


PAYL \cite{WangEtAl:2005} uses 1-grams and unsupervised learning to build a
byte-frequency distribution model of network traffic payloads. A 1-gram is simply a
single byte with value in the range 0-255. By preprocessing a packet
payload in this way the technique builds a feature vector containing the relative frequency count of
each of the 256 possible 1-grams in the payload. The model also includes the
average frequency, as well as the variance and standard deviation as other
features. Separate models of normal traffic are created for each combination of
destination port and length of the flow. During the detection phase a simplified Mahalanobis distance
measure is used to compare the current traffic to the model, and if the distance 
exceeds a given threshold PAYL raises an alert. Testing was performed on all
attacks in the DARPA 1999 dataset using individual packets and connections as data units.
The overall detection rate was close to 60\% at a false positive rate less then 1\%.

POSEIDON \cite{bolzoni2005poseidon} is based on PAYL for the detection process, while the
preprocessing step is different. Unlike PAYL, the output of a SOM classifier is used for determining 
whether to create a separate model. The aim of the SOM is to identify similar payloads for a
given destination address and port. This improvement was shown to produce less
models and higher accuracy than PAYL.

ANAGRAM \cite{WangEtAl:2006} is also based on PAYL, but uses a mixture of
high-order \textit{n}-grams with \textit{n} $>$ 1. This characteristic
makes the system less susceptible to mimicry attacks since higher order 
\textit{n}-grams are harder to emulate in padded bytes. By
contrast, it is possible to evade very easily PAYL if normal byte frequencies are known, 
since malicious payloads can be padded with bytes to match it. ANAGRAM
uses supervised learning to model normal traffic by storing legitimate \textit{n}-grams 
into one bloom filter, and models attack traffic by storing \textit{n}-grams from
malicious packets into a separate bloom filter. At runtime the \textit{n}-grams from
incoming payloads are compared with those stored in the two filters. 
If the \textit{n}-grams either match the attack  filter, or do
not match the normal  filter, the system raises an anomaly. 

Similarly, McPAD \cite{perdisci2009mcpad} creates 2\textit{v}-grams and 
constructs different features spaces, which are handled by a different classifier.
A clustering algorithm is used to reduce dimensionality of the future space, which
is $256^2 = 65536$. Multiple one-class SVMs are used for classification: their output
are combined by a meta-classifier  into a final classification prediction. 
The results of testing McPAD showed it could detect shellcode attacks in HTTP requests. 

In \cite{rieck2007language}, the authors model legitimate contents with high-order 
\textit{n}-grams, which are extracted from connection payloads.
Unsupervised anomaly detection is performed on the application layer protocol bytes.
Their approach differs from others because each high-order \textit{n}-gram
 is associated with a geometric representation. \textit{n}-grams and words in connection
payloads are compared using vectorial similarity measures such as distance
functions.

In this paper an anomaly-based intrusion detection technique is proposed, called
Packet Chunk Anomaly Detector (PCkAD), which uses \textit{n}-grams and a novel
preprocessing step for the analysis of network packets payload. The system uses
protocol knowledge to identify the relevant parts of a packet payload for the
analysis and discard all the rest, subsequently the useful payload is split up
in not overlapping portions, here called chunks, of equal length. The system
models normal behavior of clear and structured network contents in an automatic
and semi-supervised fashion. A packet payload is recognized as anomalous if it
contains too many never seen \textit{n}-grams or known \textit{n}-grams whose distribution differs
considerably from that observed during the training phase.
The experiments were conducted on the dataset DARPA 1999 to test the
effectiveness of the proposed technique. The system is able to get a detection
rate of 100\% with a false-positive rate less than 1\%, for the FTP
traffic.\newline
The rest of the paper is organized as follows. Section 2 details the proposed
technique. Section 3 presents the results and evaluations of the method applied
to the DARPA 1999 dataset. Section 4 concludes the paper.

\section{Description of the technique}
This section describes the proposed technique, called PCkAD, in detail. PCkAD
uses \textit{n}-grams (briefly described in subsection 2.1) and a novel preprocessing step for the
analysis of network packets payload (subsection 2.2). Subsequently \textit{n}-grams are extracted from the contents of the
network traffic (subsection 2.3) and a model of normal behavior is built (subsection 2.4) then
the resulting model is used to identify anomalous contents (subsection 2.5).

\subsection{The \textit{n}-gram technique}
\textit{n}-grams have been used previously in fields like
information retrieval \cite{adams1992study} and statistical natural language processing \cite{brown1992class}. With this
technique it is possible to extract sequences of symbols from a given input flow
by using a sliding window of length $n$. At each position a sequence of length $n$
is considered. Formally, the set \textit{S} of features correspond to all
possible sequences of length $n$ and is defined by:
\begin{equation}
S := \{0,..., 255\}^n.
\end{equation}
To show how the technique works, consider the artificial payload \textit{x} =
``\texttt{ooddod}'' where the set of all possible symbols is restricted to ``\texttt{o}'' and
``d''. If $n$ = 2, the sequences that can be extracted are ``\texttt{oo}'', ``\texttt{od}'',
``\texttt{dd}'', ``\texttt{do}'', and ``\texttt{od}'', respectively.

The use of \textit{n}-grams does not require expert domain knowledge to construct
relevant features, since a model of normality can be built 
in an automatic fashion from the
\textit{n}-grams occurring in a packet payload.

\begin{figure}[t] 
\centering
\fbox{\includegraphics[width=230pt]{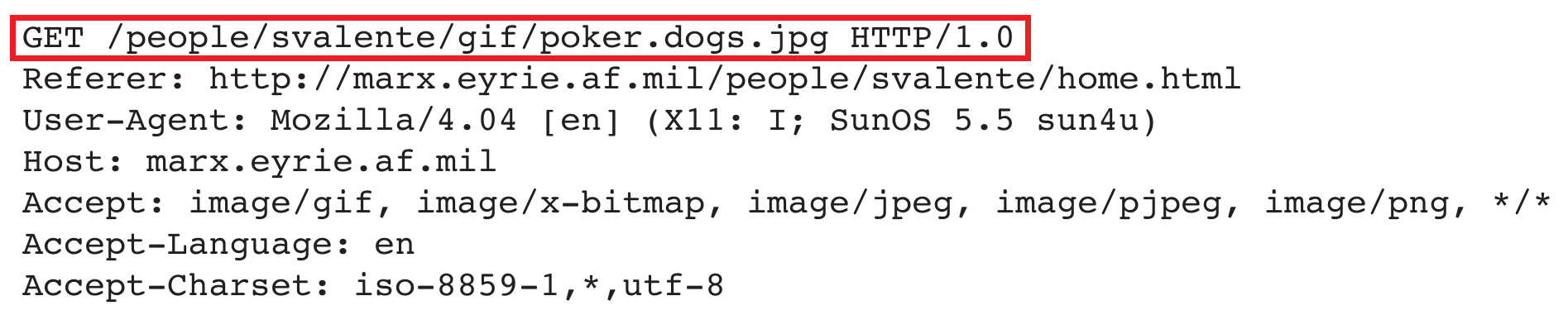}}
\caption{An example of HTTP packet payload. Only the URL following the keyword \texttt{GET}
is recognised as relevant content.}
\label{fig:http_payload_example}
\end{figure}

\subsection{The preprocessing phase}

Every time a new packet payload is observed, the system exploits protocol knowledge
to identify the relevant parts of the payload for the analysis, while
all the rest is discarded. Subsequently the useful payload is split up in non-overlapping
portions of equal length, here referred to as \textit{chunks}. More in the detail, let $P_p$ denote a packet payload and let $St_{pp}$ 
denote the set of all the basic components
of the payload structure. 

For instance, the basic components of the HTTP packet
payload are shown in Figure \ref{fig:http_payload_example}, 
and correspond to the lines starting 
with following keywords: {\tt GET}, {\tt Referer}, {\tt User-agent},
{\tt Host}, {\tt Accept}, {\tt AcceptLanguage} and 
{\tt Accept-Charset}. 

By using protocol knowledge, the IDS selects the set $Sv_{pp} \subseteq St_{pp}$ 
of the components 
which are relevant for the analysis. 
Every component $c_p$ in $Sv_{pp}$ is then split up in non-overlapping
sequences of length $len_{ck}$, in bytes,
and the
set $Ck_{cp}$ is derived. The number of chunks $nck$ that a component can contain
is defined by:
\begin{equation}
nck = \lceil len_p / len_{ck} \rceil,
\end{equation}
with $len_p$ the length of the component.

Consider again the HTTP packet of Figure \ref{fig:http_payload_example}, 
which corresponds to a GET request. 
The system recognises the URL following the keyword GET
as the only relevant component (GET and HTTP 1.0 are also included). 
If the length of a single chunk is $15$ bytes, 
then $Ck_{cp}$ will contain the following chunks: 
``{\tt GET /people/sva}'', ``{\tt lente/gif/poker}'', ``{\tt .dogs.jpg HTTP/}'',
and
``{\tt 1.0$\backslash$r$\backslash$n}''.

\subsection{The extraction phase}

There exist many IDSs which use the \textit{n}-gram technique to analyze network packets
payload, two examples are \cite{WangEtAl:2006} \cite{bolzoni2005poseidon}. 
The \textit{n}-grams are used to model the language
which characterizes a network traffic profile,
since each different \textit{n}-gram is interpreted as a
different feature of a feature space used to represent the traffic.
PCkAD extracts \textit{n}-grams from a packet payload after the
preprocessing phase described in the previous subsection.
%
%
%
The packet payload is split in chunks to learn the typical structure of a
legitimate payload. 
By exploiting the partitioning in chunks,
it is possible for the system to know which are the
typical \textit{n}-grams of a legitimate network traffic profile, how they are
distributed, and where they are typically located inside the payload. 
Only \textit{n}-grams which occur in at least a chunk are 
taken into account.
Given a chunk $c$ and a \textit{n}-gram $s$, 
$s$ is said to \textit{occur} in $c$ if:
\begin{itemize}
\item 
either $s$ is a subsequence of $c$, or
\item 
a suffix (prefix, resp.) of $s$ is a prefix (suffix, resp.)
of $c$ and the remaining part of $s$
is a suffix (prefix, resp.) of the chunk preceding (succeeding, resp.)
$c$.
\end{itemize}
The latter condition serves the purpose of taking into account 
also \textit{n}-grams located on the border between two consecutive chunks.

\begin{figure}[t] 
\centering
\fbox{\includegraphics[width=240pt]{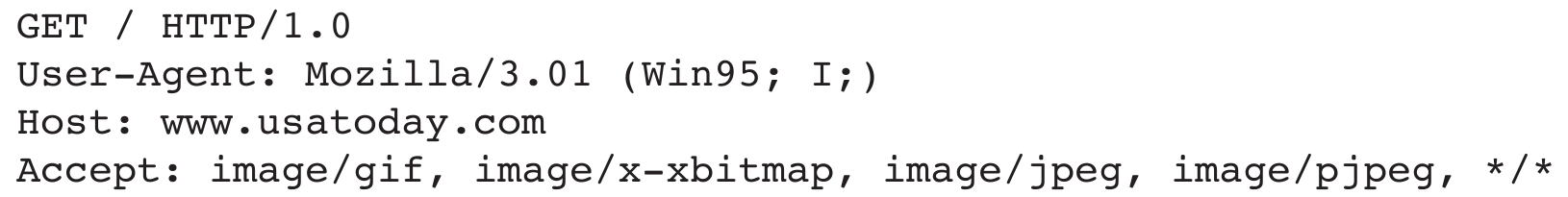}}
\caption{A new HTTP packet payload is observed.}
\label{fig:ex_step1}
\end{figure}

\begin{figure}[t] 
\centering
\fbox{\includegraphics[width=150pt]{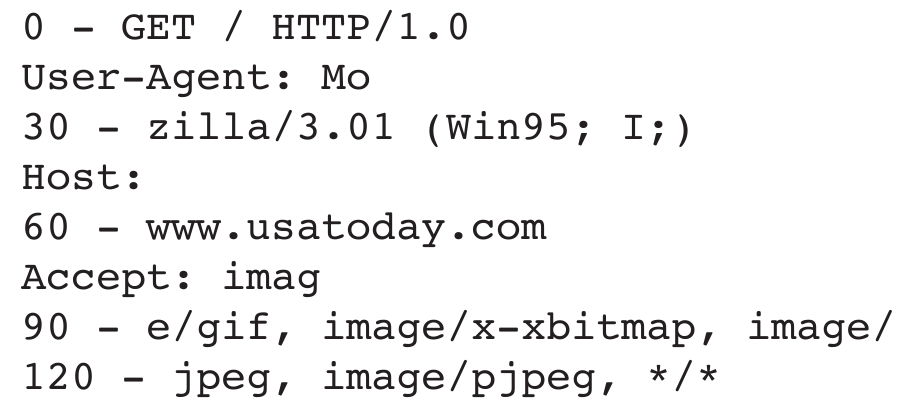}}
\caption{The payload is split up in chunks of equal length ($30$ bytes). }
\label{fig:ex_step2}
\end{figure}

\begin{figure}[htbp]
\centering
\includegraphics[width=120pt]{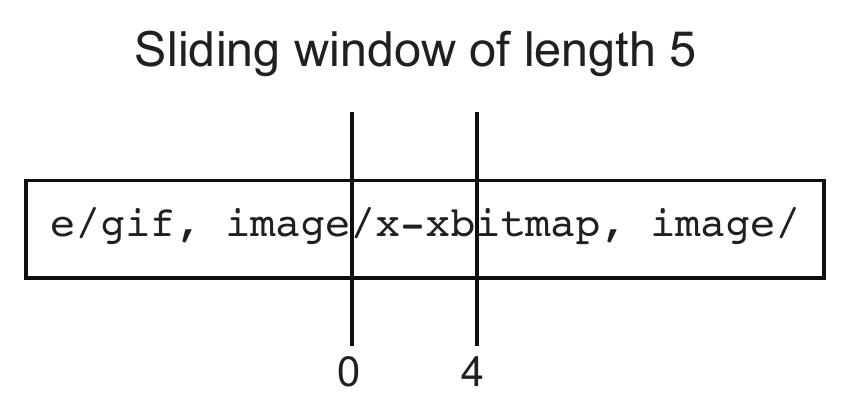}
\caption{The IDS extracts a 5-gram from a chunk.}
\label{fig:ex_step3}
\end{figure}

Figures \ref{fig:ex_step1}, \ref{fig:ex_step2} and 
\ref{fig:ex_step3} show an example of how a packet payload is processed by the IDS.
In this example, each chunk has a length of $30$ bytes and 
the entire payload is considered relevant for the analysis.
\newline


\subsection{The model}

The system models normal behavior of network traffic in an automatic and
semi-supervised fashion. 
%
In oder to build the model of normal behavior, 
an off-line learning phase working on
a training dataset containing only legitimate network packets 
is accomplished by the IDS.

Initially, the system groups the packets in classes, based on the following two 
criteria: (1) the observed port; (2) the number of chunks a packet payload contains.
Subsequently, for each different class, a model is built
that will be then exploited to assess the nature of never seen packets.

The first criterion is needed in order to 
identify network traffic profiles of different nature.
Each protocol implies specific contents, so it is important 
to separate traffic pertaining  different protocols 
to capture
information about specific characteristics of the network profile. 
Differently, the resulting classifier could exhibit a large misclassification rate.
Clearly, this strategy alone is inadequate to build accurate models. 
It is possible to observe very
different contents in the same profile. In \cite{WangEtAl:2005} it has been shown
how the byte distribution of HTTP packets vary among different length payload.

The second criterion is used to build a model from packets with similar
payload, therefore the chunks are used not only in the analysis phase
but also in the learning phase.  The chunk length  influences considerably the building process,
in that the larger the size of a chunk, the smaller the number of models. 


During the learning phase, for each built model and for each observed \textit{n}-gram, 
the average number and the standard deviation of occurrences 
are computed and stored, both for the entire relevant payload and for every single chunk.
The next section explains how these information are used in the analysis process.

\subsection{Anomalous packet detection}

Many methods have been employed in order to check if a packet is anomalous or not.
Some techniques of intrusion detection use a vectorial representation
of the packets and the achieved models then compute an anomaly index or score
by using a distance function \cite{rieck2007language}. 
The \textit{n}-gram representation fit well
with this approach: each dimension corresponds to a distinct \textit{n}-gram and its
value may be the frequency. 
However, because of the introduction of chunks, this approach cannot be directly applied in the proposed technique: \textit{n}-grams could occur in two or more chunks 
and for each of these chunks different statistics (mean and standard deviation) 
are available. Thus, the adopted anomaly index is the percentage of anomalous
\textit{n}-grams recognized in the relevant packet payload which provides
a quantitative measure of the size of anomalous portion of the payload.

As an example, consider a HTTP GET request whose relevant portion is the URL which
corresponds to the \texttt{GET} keyword. If the total number of \textit{n}-grams is 100, and 45
of them are recognized as anomalous, then the anomaly score evaluates to $45\%$.

The simplified Mahalanobis distance
is exploited to evaluate if a known \textit{n}-gram is
\textit{unusual} or \textit{non-legitimate} within the entire payload or a single chunk. 
In this work, we used the simplified assumption that
the bytes are statistically independent, so the Mahalanobis distance 
between the feature vector of a specific payload
and the model associated with the same payload
is defined by:
\begin{equation}
d(\vec{x},\vec{M}) = \sum_{i=0}^{N-1} \frac{\left | \mu_i - x_i \right |}{\sigma_i},
\end{equation}
where $N$ is the size of the set of all the possible \textit{n}-grams, 
$\mu_i$ is the average number of occurrences of the $i$-th \textit{n}-gram
in the model (aka legitimate profile) $\vec{M}$ and 
$\sigma_i$ is its standard deviation, while
$x_i$ is the count of the $i$-th \textit{n}-gram in the feature vector $\vec{x}$
associated with the 
observed packet payload.

Notice that, if an \textit{n}-gram never appears in the training samples or it appears
with exactly the same frequency in each sample, the 
standard deviation $\sigma_i$ evaluates to zero.
To avoid the distance to become infinite, a
smoothing factor $\alpha$ is added up to the standard deviation,
leading to the following formula:
\begin{equation}
d(\vec{x},\vec{M}) = \sum_{i=0}^{N-1} \frac{\left | \mu_i - x_i \right |}{\sigma_i + \alpha}.
\end{equation}
Intuitively,
the smoothing factor $\alpha$ gives information about the statistical confidence of the sampled
training data.
The confidence the samples are truly representative
of the actual distribution is inversely proportional to  $\alpha$, so for large values
of $\alpha$ the byte distribution can be more variable.


Here the Mahalanobis distance is not fully used,
but rather its terms are exploited separately in order to decide if 
an \textit{n}-gram is unusual or not. In particular, consider $v_{np}$ and $v_{nc}$ as the feature vectors
representing the normal profile for the entire relevant payload and for each chunk $c$
respectively and $v_{p}$ and $v_{c}$ as the feature vectors of a new packet payload
observed for the entire payload and for each chunk $c$ respectively. When a new
packet is observed the system uses the Mahalanobis distance to compute the
difference between each component of $v_{np}$ and $v_{p}$, for the $i$-th $n$-gram:
\begin{equation}
d (v_{np} , v_{p})_i =\frac{\left | \mu_i - x_i \right |}{\sigma_i + \alpha}
\end{equation}
the term $d (v_{np}, v_{p})_i$ is then compared with a threshold $th_s$:
if $d (v_{np}, v_{p})_i > th_s$ the $n$-gram is
said to be
\textit{unusual} (aka, \textit{non-legitimate}), otherwise it is \textit{usual} 
(aka, \textit{legitimate}).
The same approach holds for $v_{nc}$ and $v_{c}$. An $n$-gram is recognized as \textit{anomalous} if one of the following conditions holds:
\begin{itemize}
\item \textit{it has never been observed in the normal traffic}: 
in this case all its occurrences are identified as anomalous;

\item \textit{it has been observed in the normal traffic, but it 
is unusual in the entire relevant payload}:
in this case all its occurrences are identified as anomalous;

\item \textit{it has been observed in the normal traffic and 
it is usual in the entire relevant payload, but it is anyway
unusual in at least a chunk}: 
in this case only the
occurrences unusually distributed are considered anomalous.
\end{itemize}

For each packet payload the system computes:
the total number of $n$-grams ($tot_{seqs}$); %
the number of anomalous \textit{n}-grams ($a_{seqs})$.
Afterward an \textit{anomalous score}, $a_{score}$, is computed:
\begin{equation}
a_{score}= \frac{a_{seqs}}{tot_{seqs}}\cdot 100
\end{equation}
 A different threshold has been defined for each analyzed protocol.
If there is a low degree of variability in the contents of the network traffic of a specific protocol
it is appropriate to use a low threshold, while on the contrary a higher threshold is
needed to tolerate a reasonable level of variability.

\section{Experimental results}

In this section a set of experiments designed
to test
the effectiveness of the proposed technique is presented.

Experiments
were conducted on the dataset DARPA 1999. 
The 1999 DARPA IDS \cite{ideval:index.html}
data set was collected at MIT Lincoln Labs to evaluate intrusion detection systems.
All the network traffic was recorded in \texttt{tcpdump}
format and provided for evaluation. Additionally, there are also audit logs, daily
file system dumps, and BSM (Solaris system call) logs. The data consists of three weeks
of training data and two weeks of test data. The training data contains two weeks of attack-free data
and one week of data with labeled attacks. 
This dataset has been used in many
research efforts and results of tests involving this data have been reported in many publications.
Although there are problems due to the nature of the simulation environment that created
the data \cite{mchugh2000testing}, it still remains a useful set of data to compare techniques.
In \cite{lippmann2000analysis} the best results are reported.

In the experiments only the inside HTTP and FTP network traffic data which was captured between
the router and the victims was used. It is important to notice that not all the types of
payloads are suitable to be analyzed with \textit{n}-grams. If \textit{n}-grams are blindly constructed
from all packet payloads including unstructured and, in particular, encrypted data, then a huge range
of \textit{n}-grams would be created and the resulting model would not be able to distinguish
between normal and anomalous traffic in a satisfactory way. 
Therefore only clear and structured payloads are considered. The inbound TCP traffic to the ports
\texttt{80} and \texttt{21} of the hosts \texttt{172.016.xxx.xxx} was examined, since it contains most of the victims.
Each packet in the dataset was used as the data unit. The configuration set up is similar to that chosen to evaluate PAYL \cite{WangEtAl:2005}
and POSEIDON \cite{bolzoni2005poseidon}, two other intrusion detection techniques,
so that a comparison with PCkAD could be done. 
The system was trained on the DARPA dataset using week 1 (5 days, attack free)
and week 3 (7 days, attack free). The detector was then evaluated on weeks 4 and 5, which
contain 201 instances of 58 different attacks, 177 of which are visible in the inside tcpdump
data. 
In the experiments only the HTTP and FTP traffic
is considered, so the attacks using protocols TCP, UDP, ICMP, ARP
(address resolution protocol) and IP are not treated here. 


Traffic flows directed to  different ports have different byte variability, e.g.
the HTTP  payloads (port
80) are usually less variable than those of SMTP (port 25).
Hence, different thresholds were set for each protocol, in particular, their values has been derived
empirically so that the false-positive rate would be less than 1\%.
It was set a threshold of
40\% for FTP and a threshold of 30\% for HTTP. 
Other configuration parameters are:
the smoothing factor $\alpha$: it was set to 0.1, a reasonable value for this dataset;
the length of a single chunk, in bytes;
the value of \textit{n}, for the \textit{n}-grams.
The experiments were set up to assess how the chunks and the \textit{n}-grams affect the system performance.
The reported results concern only the analysis of the FTP traffic because most of the attacks
directed towards HTTP-based application where recognized because of the following reasons:
\begin{itemize}
\item the IDS identifies malformed HTTP requests easily by using protocol knowledge, like
the \texttt{crashiis} attack which sends ``{\tt GET ../..}'' as a request;
\item there exist many attack packets whose characteristics are very different from
those observed in the legitimate packets of the training data set, so it does not exist
a model for the packet classification. This means that the system automatically classify
these packets as anomalous. An example of such attack is {\tt apache2}, a DoS attack, which
sends requests with a lot of repeated ``{\tt User-Agent:sioux$\backslash$r$\backslash$n}''.
\end{itemize}
During the analysis of the FTP traffic, about 3000 FTP packets sent by the same source host,
which is sending FTP commands in a way that is typical of the Telnet protocol (one character
per packet, with the TCP flag PUSH set), were found. Although it is normal traffic, this behaviour
was never observed in the training data set. 
Currently the system is not able to handle payloads
containing only one character, because it uses high order \textit{n}-grams. Therefore these packets were
not taken into account to assess the effectiveness of PCkAD. The treatment of this aspect is
planned for future work. 
The same consideration holds for POSEIDON too \cite{bolzoni2005poseidon},
while PAYL \cite{WangEtAl:2005} works with 1-grams so it should be able to handle this scenario.
%

The variant of the technique where chunks are disabled is used as baseline for the evaluation.
Table 1 reports the results about the influence of the length of a single chunk on the analysis
process, with $n$ = 3.  $len_{ck}$,  $dr_{ck}$, $dr_w$, $fpr_{ck}$ and $fpr_w$ denote
respectively: the length of a chunk (in bytes), the detection rate with chunks enabled, the
detection rate with chunks disabled, the false-positive rate with chunks enabled and the false-positive
rate with chunks disabled.

\begin{table} [htbp]
    \caption{Results of \textit{n}-grams analysis of FTP network traffic with and without chunks.}
\centering
    \begin{tabular}{| l | l | l | l | l |}
    \hline
    $len_{ck}$ & $dr_{ck}$ & $dr_{w}$ & $fpr_{ck}$ &  $fpr_{w}$ \\ \hline
    7 &  84.2\% & 78.9\% & 0.691\% & 0.596\%\\ \hline
    15 & 100\% & 73.7\% & 0.588\% & 0.555\%\\ \hline
    20 & 84.2\% & 84.2\% & 0.518\% & 0.518\%\\ \hline
    25 & 73.7\% & 73.7\% & 0.526\% & 0.498\%\\ \hline
    39 & 63.2\% & 63.2\% & 0.358\% & 0.358\%\\ \hline
    \end{tabular}

\end{table}

As can be seen in the table, sometimes the chunks have a positive influence on
the analysis process, although typically the system produces a greater number of
false-positive when they are enabled, however this increase is very small so the
results can be considered positive.
Without chunks, with $len_{ck}$ = 7,
the system is not able to recognize properly a specific attack instance so the detection
rate goes down slightly. With $len_{ck}$ = 15 the detection rate increases from
$73,7\%$ to $100\%$ using chunks, while the false-positive rate goes up imperceptibly.
In the remaining configurations the chunks affect the results a little or nothing at all.
It may happen that the system is able to recognize anomalous packets only due to never
seen \textit{n}-grams, in this scenario the chunks have a very low influence.

The effectiveness of the chunks depends on the nature of the models, so the length
of a chunk has a great importance in the training phase. Too low values lead to
the production of many models, this means that a lot of packets with similar payloads
are split up in different classes. On the other hand, too high values would have the
opposite effect, many packets with different payloads would be grouped together.
Another experiment was set up to assess if the \textit{n}-grams are able to influence the analysis
process as well. If the value is too low then the set of all possible \textit{n}-grams the system
can learn is relatively small, so it would be harder to recognize attack instances.
With high values it is easier for the system to recognize anomalous packets, but it is
also easier to produce a greater number of false-positive. In the previous experiment
only 3-grams have been used; next 2-grams and 5-grams will be evaluated.
Tables 2, 3 and 4 show the results with 2-grams and 5-grams.

\begin{table} [htbp]
    \caption{2-grams analysis of FTP network traffic ($len_{ck}$ = 20).}
\centering
    \begin{tabular}{| l | l | l | l | l |}
    \hline
    threshold & $dr_{ck}$ & $dr_{w}$ & $fpr_{ck}$ &  $fpr_{w}$ \\ \hline
    40\% & 10.56\% & 10.56\% & 0.39\% &  0.35\%\\ \hline
    30\% & 36.84\% & 36.84\% & 1.03\% & 0.58\%\\ \hline
    25\% & 57.89\% & 57.89\% & 1.77\% & 0.79\%\\ \hline
    15\% & 78.9\% & 78.9\% & 8.02\% & 4.56\%\\ \hline
    \end{tabular}

\end{table}
\begin{table} [htbp]
    \caption{ 2-grams analysis of FTP network traffic ($len_{ck}$ = 15).}
\centering
    \begin{tabular}{| l | l | l | l | l |}
    \hline
    threshold & $dr_{ck}$ & $dr_{w}$ & $fpr_{ck}$ &  $fpr_{w}$ \\ \hline
    30\% & 42.1\% & 36.8\% & 0.67\% &  0.59\%\\ \hline
    25\% & 78.9\% & 68.4\% & 1.04\% & 0.83\%\\ \hline
    15\% & 94.7\% & 78.9\% & 6.02\% & 4.63\%\\ \hline
    \end{tabular}

\end{table}
\begin{table} [htbp]
\caption{ 5-grams analysis of FTP network traffic ($len_{ck}$ = 15).}
\centering
    \begin{tabular}{| l | l | l | l | l |}
    \hline
    threshold & $dr_{ck}$ & $dr_{w}$ & $fpr_{ck}$ &  $fpr_{w}$ \\ \hline
    40\% & 100\% & 100\% & 1.27\% &  1.23\%\\ \hline
    50\% & 100\% & 100\% & 1.07\% & 1.03\%\\ \hline
    60\% & 100\% & 100\% & 0.93\% & 0.89\%\\ \hline
    \end{tabular}
    
\end{table}

When $n$ is set to $2$ it is difficult for the system to detect attack instances, while for $n = 5$
the detection rate is very high, but the false-positive rate also goes up. As table 2 and 3 show, with 2-grams
 high false-positive rate can be observed, especially with threshold 15\%. By taking into account
 the low discriminating power of 2-grams, the resulting models are not able to classify never seen packets properly.
These results
confirm the previous considerations. In the light of the results observed, it is reasonable to say that
greater values of $n$ would lead to higher false-positive
rate while with 1-grams it would be even more difficult for the system to recognise
attack instances. In the first case it would be required a large amount of resources for
system administrators to check the alarms generated from the IDS, instead in the second
case the false-negative rate would goes up.
Table 5 compares the PCkAD's performance with those of PAYL and POSEIDON, with the false positive rate
lower than 1\% for all the techniques.
PAYL and POSEIDON are described in more details in the introduction. It is interesting
to note that PCkAD is able to get a detection rate of 100\%, like POSEIDON and better than
PAYL and it is able to get the smallest false-positive rate among the three techniques.

\begin{table} [htbp]
 \caption{Comparison among PCkAD (n = 3, $len_{ck}$ = 15), PAYL and POSEIDON; DR = detection rate, FPR = false-positive rate.}
\centering
    \begin{tabular}{| l | l | l | l | l |}
    \hline
    & PAYL & POSEIDON & PCkAD\\ \hline
    DR & 94.7\% & 100\% & 100\%\\ \hline
    FPR & 0.93\% & 0.93\% & 0.588\%\\ \hline
    \end{tabular}
   
\end{table}

\section{Conclusions and Future work}

In this paper it was presented an anomaly-based intrusion detection technique,
called Packet Chunk Anomaly Detector (PCkAD), which uses n-grams and a novel
preprocessing step for the analysis of network packets payload. The system
models the normal behaviour of network traffic profile, in an automatic and
semi-supervised fashion. The resulting models are used to classify unseen
packets.
Two experiments were set up to assess the effectiveness of the proposed
technique and the DARPA 1999 dataset was used. The system is able to get a
detection rate of 100\% with a false-positive rate less than 1\%, for the FTP
traffic. A comparison among PCkAD, PAYL and POSEIDON is reported; PCkAD and
POSEIDON share the same detection rate, which is the highest, moreover PCkAD has
got the lowest false-positive rate.

The PCkAD technique can be improved by trying %
to decrease even more the false-positive rate and 
its robustness can  be better assessed by considering other experimental scenarios.
Another relevant direction of research is being currently taking into
account is the handling of concept drift by means of weighted ensembles
of classifiers exploiting the basic strategy here presented.

\newcommand{\BIBdecl}{\setlength{\itemsep}{0.25 em}}
\bibliographystyle{IEEEtran}
\bibliography{IEEEabrv,ictai-2015}

\end{document}